\documentclass[prl,aps,twocolumn,superscriptaddress,nofootinbib,longbibliography]{revtex4-2}
\usepackage{graphicx} 
\usepackage{dcolumn}  
\usepackage{bm}       
\usepackage{amsmath}
\usepackage{epsfig}
\usepackage{color}

\begin{document}
\title{Control of the Radiative Heat Transfer in a Pair of Rotating Nanostructures}
\author{Juan R. Deop-Ruano}
\affiliation{Instituto de \'Optica (IO-CSIC), Consejo Superior de Investigaciones Cient\'ificas, 28006 Madrid, Spain}
\author{Alejandro Manjavacas}
\email{a.manjavacas@csic.es}
\affiliation{Instituto de \'Optica (IO-CSIC), Consejo Superior de Investigaciones Cient\'ificas, 28006 Madrid, Spain}
\affiliation{Department of Physics and Astronomy, University of New Mexico, Albuquerque, New Mexico 87106, USA}

\date{\today}

\begin{abstract}
The fluctuations of the electromagnetic field are at the origin of the near-field radiative heat transfer between nanostructures, as well as the Casimir forces and torques that they exert on each other. Here, working within the formalism of fluctuational electrodynamics, we investigate the simultaneous transfer of energy and angular momentum in a pair of rotating nanostructures. We demonstrate that, due to the rotation of the nanostructures, the radiative heat transfer between them can be increased, decreased, or even reversed with respect to the transfer that occurs in  absence of rotation, which is solely determined by the difference in the temperature of the nanostructures. This work unravels the unintuitive phenomena arising from the simultaneous transfer of energy and angular momentum in pairs of rotating nanostructures.
\end{abstract}

\maketitle

Radiative heat transfer between material structures originates from the thermal fluctuations of the electromagnetic field \cite{R1965}. When the distance between the structures is much smaller than the wavelength of the thermal radiation, the radiative heat transfer can greatly surpass the predictions of Planck's law due to the contribution of the near-field components of the electromagnetic field \cite{NSH09,RSJ09,SGZ14,KSF15,BMF16,CG18}. If the dimensions of the structures also fall within that range, the strong responses produced by their electromagnetic resonances provide a further enhancement of the radiative heat transfer \cite{DVJ05,NC08,PRL08,DK10,ama18,RSM17,ama70,BMV21}. 

Another important phenomenon originating from the vacuum and thermal fluctuations of the electromagnetic field is Casimir interactions \cite{C1948,DMR11,WDT16,RIB22}. 
These interactions produce forces and torques between neutral objects \cite{L07,RCJ11,GGL15,ama50,SGP18,ACG20,XJG22,SGM22}, which can play an important role in the mechanical behavior of nanostructures \cite{MC10,PSS20}. For instance, the Casimir force produces a friction for two parallel surfaces in relative motion as well as for an atom moving near a surface \cite{P97,S14,KBD17}.  Analogously, the Casimir torque acting on a rotating nanostructure generates a friction that opposes the rotation and eventually stops it \cite{ama7,ama9,ama19,BL15,XL17}. For systems containing multiple nanostructures, the Casimir torque enables the transfer of angular momentum between them \cite{RMP17,AE17,ama66}. 
Furthermore, these phenomena can even result in an analog to the Sagnac effect \cite{MSN21}.

Radiative heat transfer and Casimir interactions are usually investigated separately. The former is typically studied for ensembles of motionless nanostructures, while, for the latter, it is common to assume that all of the nanostructures are at the same temperature. However, as we show in this work, the interplay between radiative heat transfer and Casimir interactions can give rise to very interesting phenomena.

In this Letter, we characterize the transfer of energy and angular momentum in a pair of rotating nanostructures with different temperatures and rotation frequencies. Thanks to the simultaneous study of both phenomena, we demonstrate that the rotation of the nanostructures can significantly modify the radiative heat transfer between them. In absence of rotation, the energy transfer is determined by the difference in the temperature of the nanostructures, and is always directed from the hot nanostructure to the cold one \cite{CLV08,DK10,ama18}. However, when the nanostructures rotate, the radiative heat transfer can be enhanced, reduced, or even reversed, \textit{i.e.}, made to go from the cold nanostructure to the hot one, by adjusting their rotation frequencies. Our results, which are based on the fluctuational electrodynamics framework \cite{PV1971,VP07} and the dipolar approximation \cite{BBJ11,N14,MTB13,DZL17,ama75}, provide the theoretical foundations to understand how the transfer of angular momentum modifies the transfer of energy in pairs of rotating nanostructures. 

\begin{figure}
\begin{center}
\includegraphics[width=70mm,angle=0]{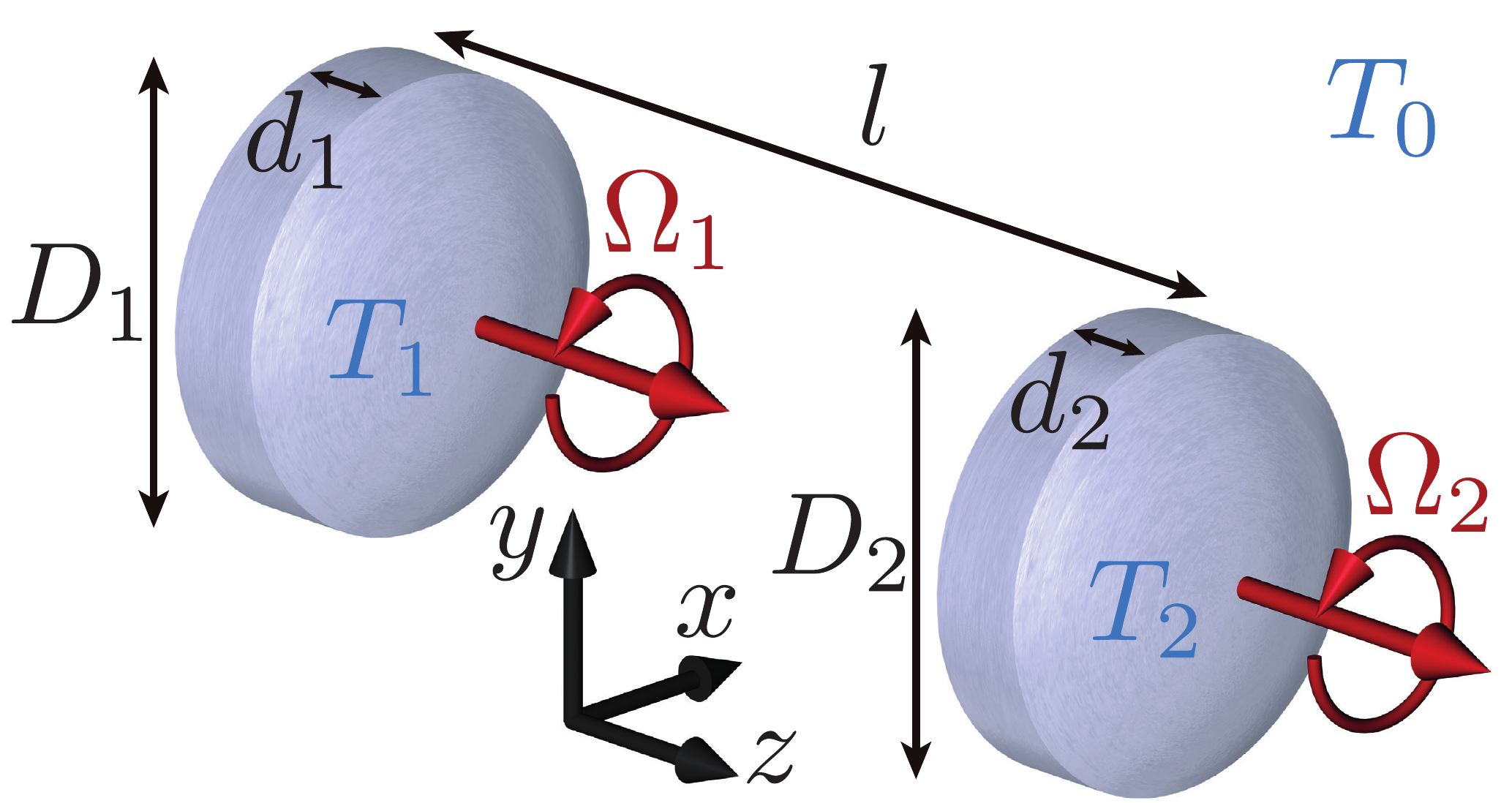}
\caption{The system under study consists of two axially symmetric nanostructures separated by a distance $l$ along the $z$ axis. The nanostructures have dimensions $D_1$, $d_1$ and $D_2$, $d_2$, are at temperatures $T_1$ and $T_2$, and rotate with frequencies $\Omega_1$ and $\Omega_2$. The temperature of the environment is $T_0$.} \label{fig1}
\end{center}
\end{figure}


The system under consideration is depicted in Figure~\ref{fig1}. It consists of two axially symmetric nanostructures separated by a distance $l$ along their symmetry axis, which we choose to be the $z$ axis. The nanostructures have dimensions $D_1$, $D_2$ and $d_1$, $d_2$ along the directions perpendicular and parallel to the $z$ axis, respectively. They rotate around the $z$ axis with rotation frequencies $\Omega_1$ and $\Omega_2$. 
The temperatures of the nanostructures, as defined in the frame at rest with each of them, are $T_1$ and $T_2$ while, for the environment, it is $T_0$.
We assume that the size of the nanostructures is significantly smaller than both their separation and the relevant wavelengths of the system, which are determined by the temperature of the nanostructures as well as their rotation and resonance frequencies. These assumptions allow us to work within the dipolar approximation, in which the nanostructures are modeled as electric point dipoles. 

We are interested in simultaneously studying the transfer of angular momentum and energy between the rotating nanostructures. To characterize these transfers, we calculate, respectively, the power radiated by the nanostructures and the electromagnetic torque acting on them. Within the dipolar limit, the torque acting on nanostructure $1$ is given by $M_1 = \langle \mathbf{p}_1(t) \times \mathbf{E}_1(t) \rangle \cdot  \mathbf{\hat{z}}$, while the power that it radiates is $P_{1} = - \langle \mathbf{E}_1(t) \cdot \partial \mathbf{p}_1(t)/\partial t  \rangle$. Here, $\mathbf{p}_1(t)$ and $\mathbf{E}_1(t)$ represent the self-consistent electric dipole and electric field in nanostructure $1$ and the brackets indicate the average over fluctuations, which we perform using the fluctuation-dissipation theorem \cite{N1928,CW1951,ama7}. Following the procedure described in the Appendix, we obtain the following expressions for the torque: 
\begin{equation}
M_1 = - \int_{-\infty}^{\infty} \!\!\!\! d\omega \left[F^+ N_1^{-} - G^+ N_2^-\right] \label{M1}
\end{equation}
and the power radiated:
\begin{equation}
P_1 = \int_{-\infty}^{\infty} \!\!\!\! d\omega \omega \left[F^+ N_1^{-} - G^+ N_2^- + F^z N^z_1 - G^z N^z_2\right] , \label{P1}
\end{equation}
where  $N_i^-=n(T_i,\omega-\Omega_i) - n(T_0,\omega)$ and $N_i^z=n(T_i,\omega)/2 - n(T_0,\omega)/2$,  with $n(T,\omega) = [\exp(\hbar \omega / k_{\rm B} T)-1]^{-1}$. 
Furthermore, the functions $F^{\nu}$ and $G^{\nu}$ are defined as $F^{\nu}=(2\hbar/\pi) |h^{\nu}|^2 {\rm Im}\{ \chi_{1}^{\nu} \} {\rm Im}\left\{g_{0}+\alpha_2^{\nu}(g^{\nu})^2\right\}$ and $G^{\nu}=(2\hbar/\pi) |h^{\nu}g^{\nu}|^2 {\rm Im}\{ \chi_{1}^{\nu} \} {\rm Im}\{\chi_2^{\nu}\}$, with $h^{\nu} = [1-\alpha_{1}^{\nu}\alpha_2^{\nu}(g^{\nu})^2]^{-1}$, $\chi_i^{\nu} = \alpha_i^{\nu}-g_0|\alpha_i^{\nu}|^2$, $g_0=2ik^3/3$, $g^{\nu}=\exp(ikl)[(1-\delta_{\nu z})k^2/l+(1-3\delta_{\nu z})(ik/l^2-1/l^3)]$, $k=\omega/c$, and $\delta_{\nu \mu}$ being the Kronecker delta. Importantly, we can obtain analogous expressions for nanostructure $2$ by interchanging the indices $1\leftrightarrow 2$ in Equations~(\ref{M1}) and (\ref{P1}).
 
In these expressions, $\alpha_i^z$ represents the component of the polarizability of the nanostructure along the $z$ axis, which is not affected by the rotation. On the other hand, $\alpha_i^{\pm}$ denotes the components of the effective polarizability seen from the frame at rest, in the basis formed by the unit vectors $\mathbf{\hat{e}}^{\pm} = (\mathbf{\hat{x}}\pm i\mathbf{\hat{y}})/ \sqrt{2}$. The calculation of these components is not trivial. 
If the intrinsic response of the nanostructure is assumed to remain unchanged by the rotation, and the Coriolis and centrifugal effects are neglected, the effective polarizability only accounts for the effect of the Doppler shift caused by the rotation \cite{ama7,ama9,ama19,ama50,MGK13,MGK13_2,MJK14,LS16}. 
However, it was later shown that the inclusion of these effects gives rise to corrections that can partially or completely cancel the effect of the Doppler shift and introduce other dependences with the rotation frequency \cite{PXG19,PXG19_2,ama66,PXG21}. 

Equations~(\ref{M1}) and (\ref{P1}) allow us to calculate the transfer of angular momentum and energy between the rotating nanostructures by numerically computing the integrals over frequency. Nevertheless, they do not provide direct insight into the physical mechanisms behind these phenomena. Here, instead, we derive closed-form analytical expressions by considering the following approximations. First, we assume that the rotation frequencies $\Omega_1$ and $\Omega_2$, as well as the thermal frequencies $\theta_1=2\pi k_{\rm B}T_1/\hbar$ and $\theta_2=2\pi k_{\rm B}T_2/\hbar$, are all much smaller than both $c/l$ and the resonance frequencies of the nanostructures $\omega_{\rm r, 1}$ and $\omega_{\rm r, 2}$.
Second, we assume that $D_1 \gg d_1$ and $D_2 \gg d_2$, which allows us to neglect the component of the polarizabilities along the $z$ axis for both nanostructures. Moreover, as shown in the Appendix, we describe $\alpha_i^{\pm}$ using a harmonic oscillator model and taking into account the Coriolis and centrifugal effects. With the approximations described above, 
the polarizability reduces to $\alpha_i^{\pm}(\omega) \approx a_i [1 + i (\omega \mp \Omega_i) \gamma_i / \omega_{{\rm r}, i}^2]$, where $\gamma_i$ is the nonradiative damping of the electromagnetic resonance, and $a_i$ is a constant with units of polarizability.  Finally, we neglect multiple scattering effects, \textit{i.e.}, $h^{\nu} \approx 1$.  Upon applying all of these approximations, the torque acting on nanostructure $1$ can be approximated by
\begin{equation}
\frac{M_1}{C} = -{\left(\Omega_{1}-\Omega_{2}\right)\left(\theta_{1}^{2}+\theta_{2}^{2}\right)- 2 \left(\Omega_{1}-\Omega_{2}\right)^{3} }, \label{M1_a}
\end{equation}
 and the power radiated by
\begin{equation}
\begin{aligned}
\frac{P_{1}}{C}  = {}& \frac{\Pi_1}{C}+ \frac{1}{2}(\theta_1^2 - \theta_2^2) (\Omega_1 - \Omega_2)^2  \\
& + \left(\Omega_1^2-\Omega_2^2\right) \left[\frac{1}{2} \left(\theta_1^2+\theta_2^2\right)+\left(\Omega_1-\Omega_2\right)^2\right]. \label{P1_a}
\end{aligned}
\end{equation}
Here,  $\Pi_{1}=C(\theta_1^4-\theta_2^4)/10$ represents the power radiated by nanostructure $1$ when neither of the nanostructures rotate,  which has the same dependence with  temperature as the Stefan-Boltzmann law. Furthermore, $C$, defined as 
\begin{equation}
C = \frac{\hbar}{6 \pi}\left(\frac{\gamma_{1} \gamma_{2}}{\omega_{\rm r, 1}^2 \omega_{\rm r, 2}^2}\right) \left(\frac{a_1 a_2}{l^6}\right), \nonumber
\end{equation}
contains all of the information of the electromagnetic response of the nanostructures, and, in particular, of their material properties. Again, analogous expressions for nanostructure $2$, \textit{i.e.}, $M_2$ and $P_2$, can be obtained by interchanging the indices $1\leftrightarrow 2$.

\begin{figure}
\begin{center}
\includegraphics[width=70mm,angle=0]{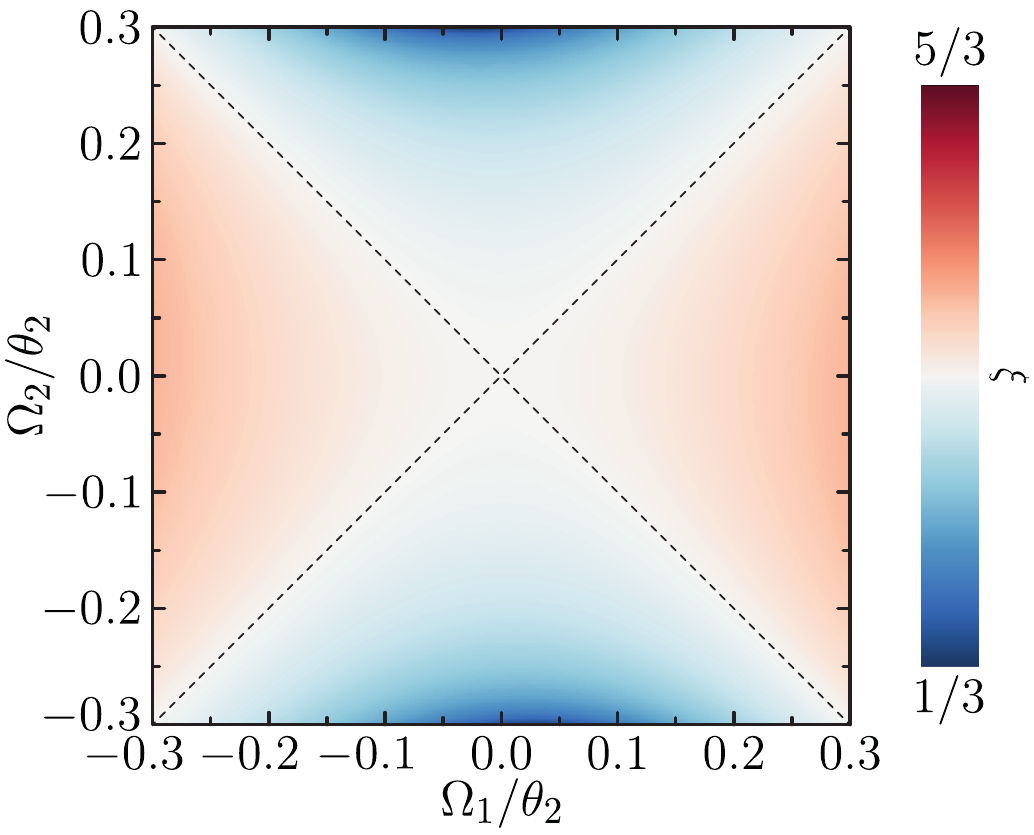}
\caption{Effective temperature ratio $\xi$ for two nanostructures rotating with different frequencies. The dashed lines indicate the values of $\Omega_1$ and $\Omega_2$ for which the ratio is equal to one.} 
\label{fig2}
\end{center}
\end{figure}

The expressions above fully describe the transfer of angular momentum and energy between the two nanostructures. Examining them, we can readily verify that $M_1+M_2=0$ and $P_1+P_2 = 0$, which tells us that, under the approximations considered above, there is no transfer of angular momentum or energy between the nanostructures and the environment. For this reason, the two rotating nanostructures behave as a closed system. This is, indeed, expected, since the transfer to the environment occurs through far-field radiation, which, with the approximations detailed above, is negligible compared with the near-field interactions that determine the transfer between the nanostructures. Therefore, in the remainder of this work, we refer to $P_1$ as the power transferred between the nanostructures. 

Equations~(\ref{M1_a}) and (\ref{P1_a}) show that the torque and the power transferred depend on both the temperatures and the rotation frequencies of the two nanostructures. 
In particular, the sign of the torque is determined by the difference between the rotation frequencies of the nanostructures, and therefore vanishes when their rotation is synchronized. On the other hand, the temperature of the nanostructures only affects the magnitude of the torque, which remains finite even for $\theta_1=\theta_2=0$, provided that $\Omega_1\neq\Omega_2$. 

The power transferred, however, shows a very different behavior: both its sign and magnitude depend on a nontrivial interplay between the rotation frequencies and the temperatures of the nanostructures. Interestingly, examining the last term of Equation~(\ref{P1_a}), we note that it is possible to obtain a nonzero $P_1$ even when the nanostructures have equal temperatures. In other words, the rotation of the nanostructures enables a transfer of energy, as if there were a temperature difference between them. To analyze this effect, we calculate the ratio between the temperatures $\xi=\theta_1/\theta_2$ that two nonrotating nanostructures need to have to produce the same power transferred as two rotating nanostructures with equal temperatures $\theta_1=\theta_2$. This ratio, which is given by
\begin{equation}
\xi =   \left\{1 + 10 \frac{\Omega_{1}^{2}-\Omega_{2}^{2}}{\theta_2^2} \left[1 + \frac{\left(\Omega_{1}-\Omega_{2}\right)^{2}}{\theta_2^2} \right] \right\}^{1/4}, \nonumber
\end{equation}
is plotted in Figure~\ref{fig2} as a function of both $\Omega_1$ and $\Omega_2$. Examining these results, we observe that, for $|\Omega_1|>|\Omega_2|$, $\xi>1$, while for, for $|\Omega_1|<|\Omega_2|$, the opposite is true.  This means that, for nanostructures with equal temperatures, the power is always transferred from the nanostructure with the larger magnitude of rotation frequency to the nanostructure with the smaller one. Furthermore, the minimum and maximum values of $\xi$ are achieved when the magnitude of the rotation frequency of one of the nanostructures is much smaller than the other. 
On the other hand, when $|\Omega_1| = |\Omega_2|$ (dashed lines), the temperature ratio is one and, therefore, the power transferred vanishes.

\begin{figure}
\begin{center}
\includegraphics[width=70mm,angle=0]{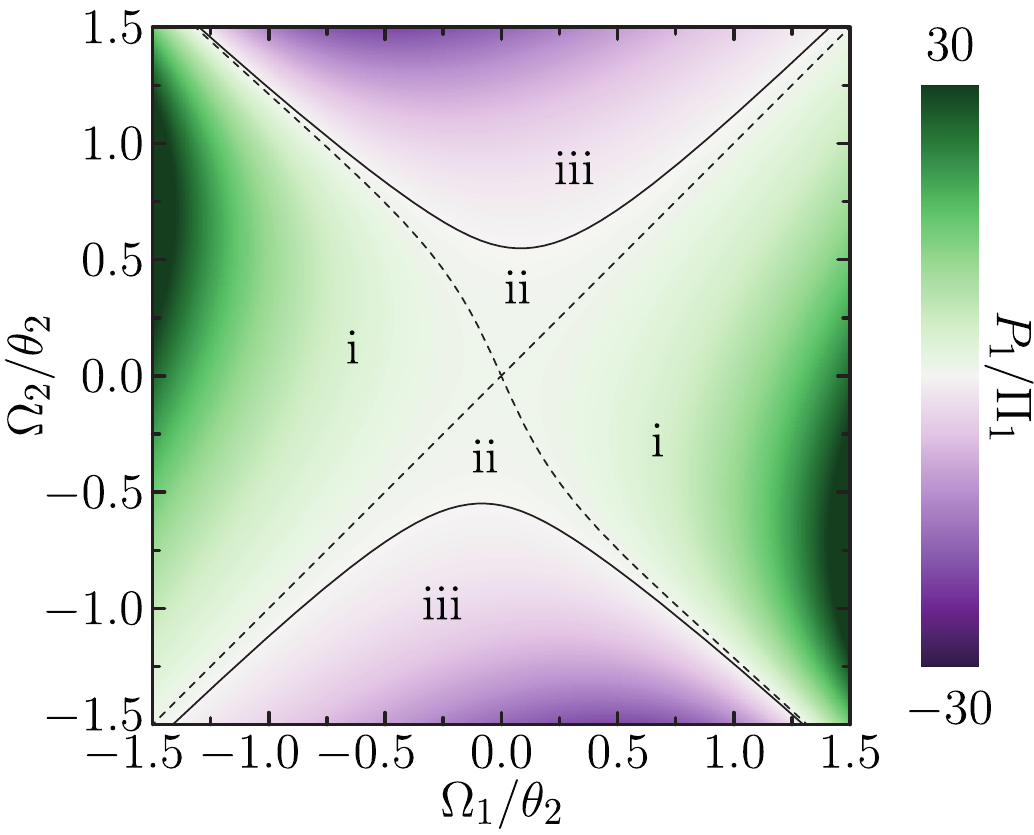}
\caption{Power transferred between two rotating nanostructures normalized to $\Pi_1$ for $\theta_1 = 1.5\theta_2$. The black solid and dashed curves signal $P_1/\Pi_1=0$ and $P_1/\Pi_1=1$, respectively, while the labels indicate the regimes in which: $P_1 / \Pi_1 > 1$ (i), $0 < P_1 / \Pi_1 < 1$ (ii), and $P_1 / \Pi_1 < 0$ (iii).} \label{fig3}
\end{center}
\end{figure}

The results of Figure~\ref{fig2} demonstrate that, for two nanostructures with equal temperatures, the rotation induces a transfer of energy and hence is effectively equivalent to a temperature difference. 
Therefore, if the nanostructures have different temperatures, we expect the rotation to modify the power transferred and even allow its direction to be changed. In order to explore this possibility, in Figure~\ref{fig3}, we plot $P_1/\Pi_1$ as a function of $\Omega_1$ and $\Omega_2$ for $\theta_1=1.5\theta_2$. We use black solid and dashed curves to indicate the frequencies for which $P_1/\Pi_1$ is equal to $0$ and $1$, respectively. These curves separate the results into three different regimes. In the first regime, $P_1/\Pi_1>1$, so the rotation serves to increase the power transferred between the nanostructures with respect to the nonrotating case. The enhancement is maximum when the nanostructures rotate in opposite directions but with the hotter nanostructure rotating faster than the colder one. The second regime is characterized by $0 < P_1 / \Pi_1 < 1$, which means that the rotation reduces the power transferred. In this case, the contribution to $P_1$ of the terms in Equation~(\ref{P1_a}) that depend on the rotation frequencies counteract $\Pi_1$, thus producing a decrease in the power transferred. When the combination of these terms surpasses $\Pi_1$ in magnitude, the power transferred changes its direction, going from the colder nanostructure to the hotter one. This corresponds to the third regime for which $P_1 / \Pi_1 < 0$. Importantly, this effect is maximized when the nanostructures rotate in opposite directions with the colder nanostructure rotating faster than the hotter one. The results of Figure~\ref{fig3} confirm that the transfer of angular momentum between the rotating nanostructures modifies the transfer of energy. Indeed, when $\Omega_1=\Omega_2$, the transfer of angular momentum vanishes (since $M_1=0$) and, expectedly, $P_1=\Pi_1$.

In order to get insight into the physical mechanisms that give rise to the different regimes illustrated in Figure~\ref{fig3}, we analyze the change in the internal energy of the nanostructures. For nanostructure $1$, this quantity is defined as $\dot{U}_1 = -P_1 -M_1 \Omega_1$, where $-M_1\Omega_1$ represents the rate of decrease of its mechanical energy. Notice that we define the internal energy as all of the energy of the nanostructure that is not mechanical. Upon insertion of Equations~(\ref{M1_a}) and (\ref{P1_a}), we obtain 
\begin{equation}
\frac{\dot{U}_1}{C} =-\frac{\Pi_1}{C} + \theta_{2}^{2}\left(\Omega_{1}-\Omega_{2}\right)^{2}+ \left(\Omega_{1}-\Omega_{2}\right)^{4}, \nonumber
\nonumber
\end{equation}
with a similar expression for nanostructure $2$ obtained by interchanging the indices $1\leftrightarrow2$. While the first term of the equation corresponds to the power radiated in absence of rotation, the rest of the terms, which are always positive, only contribute when there is a difference in the rotation frequencies. We can also compute the change in the internal energy of the entire system, which is given by
\begin{equation}
\frac{\dot{U}_1 + \dot{U}_2}{C} = (\theta_1^2 + \theta_2^2)(\Omega_1 - \Omega_2) ^2 + 2 (\Omega_1 - \Omega_2) ^4. \nonumber
\end{equation}
Interestingly, this quantity always increases, regardless of the temperatures of the nanostructures. Note that, since this is a closed system, this increase in the internal energy must come from a decrease in the mechanical energy of the system. The equations above provide us with information about the equilibrium conditions for the system. In particular, to simultaneously obtain $\dot{U}_1=0$ and $\dot{U}_2=0$, it is necessary that $\theta_1=\theta_2$ and $\Omega_1=\Omega_2$.

\begin{figure}
\begin{center}
\includegraphics[width=70mm,angle=0]{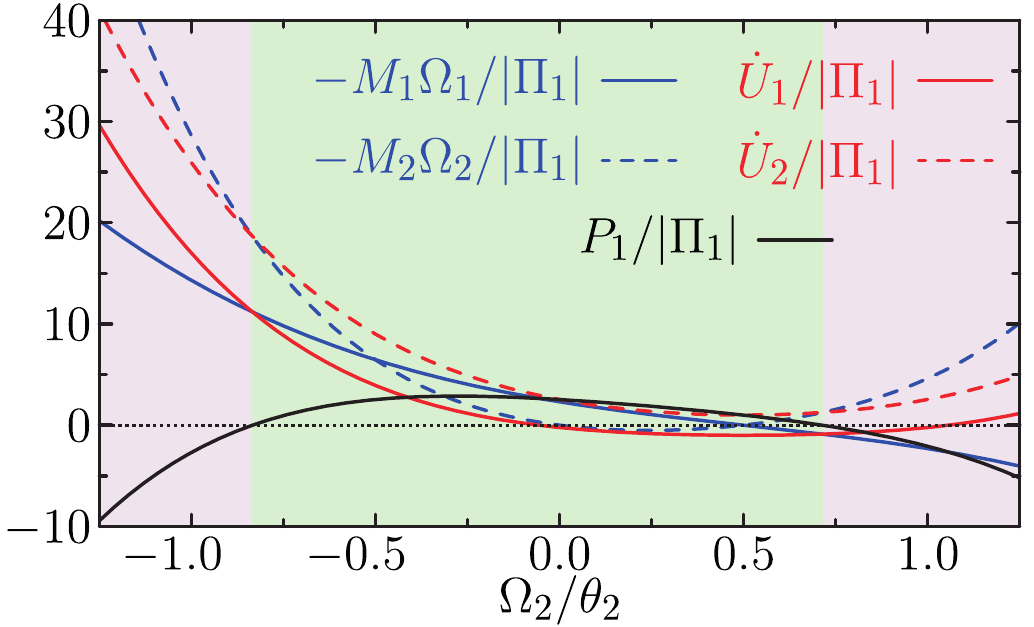}
\caption{Power transferred (black curve) and rates of change in mechanical energy (blue curves) and internal energy (red curves) for nanostructure $1$ (solid curves) and $2$ (dashed curves). All quantities are normalized to $|\Pi_1|$. We assume $\Omega_1 = 0.5 \, \theta_2$ and $\theta_1 = 1.5 \, \theta_2$. Regions of negative (positive) $P_{1}$ are indicated with a purple (green) background.}  \label{fig4}
\end{center}
\end{figure}

Figure~\ref{fig4} compares the value of the power transferred (black curve) with the rates of change of the mechanical (blue curves) and internal energy (red curves) of the nanostructures. We use solid and dashed curves for nanostructures $1$ and $2$, respectively. We assume $\Omega_1=0.5\Omega_2$ and $\theta_1=1.5\theta_2$ and plot all of the values as a function of $\Omega_2$. Unlike in Figure \ref{fig3}, here, we normalize all of the quantities to $|\Pi_1|$ so their sign is not altered.  We signal the regions in which the power transferred is reversed using a purple background, while the green background indicates that $P_1$ goes in the direction of $\Pi_1$. As expected, the boundaries between these two regions are located at the values of $\Omega_2$ for which the red and blue curves cross, since, at these points, the power transferred between the nanostructures vanishes. Furthermore, in the regions where the direction of the power transferred is reversed, the rate of decrease in mechanical energy of the colder nanostructure, $-M_2 \Omega_2$, is larger than all of the other terms analyzed. This means that, for the colder nanostructure, the decrease in mechanical energy is larger than the increase in its internal energy. For the hotter nanostructure, on the other hand, we observe the exact opposite situation, that is, $\dot{U}_1>-M_1\Omega_1$. This combination of behaviors is at the origin of the reversal of the direction of the power transferred. Indeed,  the power transferred satisfies $2P_1=\dot{U}_2-\dot{U}_1+M_2\Omega_2-M_1\Omega_1$. Therefore, in order to reach $P_1<0$ for $\theta_1 > \theta_2$, it is necessary that  $-M_2\Omega_2+M_1\Omega_1>\dot{U}_2-\dot{U}_1$. This condition is clearly satisfied in the regions with a purple background, thus confirming that the change in the direction of the power transferred is associated with the imbalance between the decrease in the mechanical energy and the change in the internal energy of the nanostructures.

In summary, we have studied the simultaneous transfer of energy and angular momentum in a pair of rotating nanostructures. To that end, working within the framework of fluctuational electrodynamics and the dipolar approximation, we have derived analytical expressions for the torque and power transferred between the nanostructures. We have shown that, for equal temperatures but different rotation frequencies, there is a power transferred from the nanostructure that rotates faster to the one rotating at a slower frequency. When there is also a difference in the temperatures of the nanostructures, the power transferred displays a rich behavior arising from the nontrivial interplay between temperature and rotation frequency. In particular, we have shown that, depending on the rotation frequency of the nanostructures, the power transferred can be enhanced or reduced with respect to that of a nonrotating pair. Furthermore, it is also possible to reverse the direction of the transfer of energy, making it go from the colder to the hotter nanostructure. 
It is worth noting that our results are derived in the limit in which the thermal and rotation frequencies are much smaller than the resonances of the nanostructures. Therefore, the behaviors described here can be enhanced by exploiting the electromagnetic resonances of the system. Moreover, although we have considered a pair of nanostructures as a canonical example, our model can be readily applied to any material structure with a dipolar resonance as, for instance, a large molecule. 
Importantly, the range of temperatures for which the effects described in this work can take place is determined by the rotation frequencies of the nanostructures. This means that for temperatures of the order of $1$K, the rotation frequencies need to be $\sim100\,$GHz. These rotation frequencies are within experimental reach for molecules like fullerenes \cite{DDE96} and are one to two orders of magnitude beyond the rotation frequencies already achieved for nanostructures \cite{RDH18,AXB18,AXB20}.
Our work provides fundamental understanding of how the transfer of angular momentum produced by the Casimir torque modifies the transfer of energy in rotating nanostructures. Therefore, the results of this work can be exploited to find new avenues to control the radiative heat transfer between nanoscale objects.

\acknowledgments
This work was sponsored by Grant No.~TEM-FLU PID2019-109502GA-I00 funded by MCIN/AEI/ 10.13039/501100011033, as well as the U.S. National Science Foundation (Grant No.~DMR-1941680). We  acknowledge support of a 2022 Leonardo Grant for Researchers in Physics, BBVA Foundation. J.R.D-R. acknowledges a predoctoral fellowship from the MCIN/AEI assigned to Grant No.~PID2019-109502GA-I00.

\onecolumngrid
\appendix
\section{Appendix}\label{ap}
\renewcommand{\thefigure}{S\arabic{figure}}
\setcounter{figure}{0} 

\subsection{Derivation of the torque and the power radiated}

As described in the main text, the system under study consists of two axially symmetric nanostructures rotating around their symmetry axis, which we choose as the $z$ axis, with frequencies $\Omega_1$ and $\Omega_2$. Both of these frequencies are defined with respect to the frame at rest. The nanostructures are separated by a distance $l$ along the $z$ axis and have dimensions $D_1$, $D_2$ and $d_1$, $d_2$ in the directions perpendicular and parallel to that axis. The temperatures of the nanostructures, as defined in the frame at rest with each of them, are $T_1$ and $T_2$, respectively, while the environment surrounding the nanostructures is at temperature $T_0$. As explained in the main text, we assume that the nanostructures are small with respect to both $l$ and the relevant wavelengths of the problem, which are determined by their temperatures as well as their rotation and resonant frequencies. This approximation allows us to model the nanostructures as rotating electric point dipoles with a response characterized by an effective polarizability. To quantify the transfer of energy and angular momentum, we calculate the power radiated by the nanostructures and the torque exerted on them. In particular, the torque exerted on nanostructure $1$ is given by $M_1 = \langle \mathbf{p}_1(t) \times \mathbf{E}_1(t) \rangle \cdot \mathbf{\hat{z}}$, where $\mathbf{p}_1(t)$ and $\mathbf{E}_1(t)$ are, respectively, the electric dipole moment and the electric field in the nanostructure and the angle brackets $\langle \rangle$ denote the average over fluctuations. In the same way, the power radiated by nanostructure $1$ can be written as $P_1 = - \langle \mathbf{E}_1(t)\cdot \partial \mathbf{p}_1(t)/\partial t \rangle$. To continue our derivation, we change to the frequency domain using the Fourier transform, defined as ${\bf p}_i(t)=\int_{-\infty}^{\infty} \frac{d\omega}{2 \pi} {\bf p}_i(\omega) e^{-i \omega t}$ for the dipole and similarly for the field. Furthermore, we work in the basis formed by the unit vectors $\hat{\mathbf{e}}^{\pm}=(\hat{\mathbf{x}} \pm i \hat{\mathbf{y}})/\sqrt{2}$ and $\hat{\mathbf{z}}$. With all of these changes,  the torque becomes
\begin{equation}
M_1= \frac{i}{4 \pi^2} \int_{-\infty}^{\infty} d \omega d\omega' e^{-i\left(\omega-\omega'\right) t} \left[\left\langle p_1^{+ *}\left(\omega'\right) E_1^{+}(\omega)\right\rangle - \left\langle p_1^{- *}\left(\omega'\right) E_1^{-}(\omega)\right\rangle\right], \label{SI_M1}
\end{equation}
while the power radiated is given by 
\begin{equation}
P_1= -\frac{i}{4 \pi^2} \int_{-\infty}^{\infty} d\omega d\omega' e^{-i\left(\omega-\omega'\right) t} \omega' \left[\left\langle p_1^{+ *}\left(\omega'\right) E_1^{+}(\omega)\right\rangle + \left\langle p_1^{- *}\left(\omega'\right) E_1^{-}(\omega)\right\rangle + \left\langle p_1^{z *}\left(\omega'\right) E_1^{z}(\omega)\right\rangle\right]. \label{SI_P1}
\end{equation}
In these equations, we use the symbol $^*$ to denote the complex conjugate. Moreover, $p_1^{\nu}(\omega)$ and $E_1^{\nu}(\omega)$ represent the different components of the self-consistent dipole and field, which are obtained by solving 
\begin{equation}
p_1^{\nu}(\omega)=p_1^{{\rm fl,} \nu}(\omega)+\alpha_{1}^{\nu}(\omega) E_1^{{\rm fl,} \nu}(\omega)+\alpha_{1}^{\nu}(\omega) g^{\nu}(\omega) p_2^{\nu}(\omega)  \nonumber
\end{equation}
for the dipole and 
\begin{equation}
E_1^{\nu}(\omega)= E_1^{{\rm fl,} \nu}(\omega)+g_0(\omega) p_1^{\nu}(\omega)+ g^{\nu}(\omega) p_2^{\nu}(\omega) \nonumber
\end{equation}
for the electric field. Here, $\alpha_i^{\nu}$ represents the different components of the effective polarizability of the nanostructures, while the function $g^{\nu}(\omega)= \exp(i k l)\left[\left(1-\delta_{\nu z}\right)k^2 / l+\left(1-3 \delta_{\nu z}\right)\left(i k / l^2-1 / l^3\right)\right]$ describes their interaction. Moreover, $\delta_{\nu \mu}$ is the Kronecker delta, $g_0(\omega) = 2i k^3 / 3$, and $k = \omega / c$. The equivalent equations for nanostructure $2$ are obtained  simply by interchanging the indices $1\leftrightarrow2$. The solution for the self-consistent dipole is given by
\begin{equation}
p_1^{\nu}(\omega) / h^{\nu}(\omega) = p_1^{{\rm fl,} \nu}(\omega)+\alpha_1^{\nu}(\omega) g^{\nu}(\omega) p_2^{{\rm fl,} \nu}(\omega) + \alpha_1^{\nu}(\omega) E_1^{{\rm fl,} \nu}(\omega) +\alpha_1^{\nu}(\omega) \alpha_2^{\nu}(\omega) g^{\nu}(\omega) E_2^{{\rm fl,} \nu}(\omega)  ,
\nonumber
\end{equation}
while, for the electric field, we get
\begin{align}
E_1^{\nu}(\omega) / h^{\nu}(\omega) = {}& \left[1+g_0(\omega) \alpha_1^{\nu}(\omega)\right] E_1^{{\rm fl,} \nu}(\omega)+ g^{\nu}(\omega)  \alpha_2^{\nu}(\omega)\left[1+g_0(\omega) \alpha_1^{\nu}(\omega)\right] E_2^{{\rm fl,} \nu}(\omega) \nonumber \\
{}& +\left[g_0(\omega) + \alpha_2^{\nu}(\omega) (g^{\nu}(\omega))^2\right] p_1^{{\rm fl,} \nu}(\omega)+g^{\nu}(\omega) \left[1+g_0(\omega) \alpha_1^{\nu}(\omega)\right] p_2^{{\rm fl,} \nu} (\omega),
\nonumber
\end{align}
where $h^{\nu}(\omega) = \left[1 - \alpha_1^{\nu}(\omega) \alpha_2^{\nu}(\omega) (g^{\nu}(\omega))^2\right]^{-1}$.
Then, we can calculate the torque and the power radiated by introducing these solutions into Equations~(\ref{SI_M1}) and (\ref{SI_P1}) and performing the average over fluctuations using the fluctuation-dissipation theorem (FDT) \cite{N1928,CW1951,ama7}. For the dipoles, we apply the FDT in the frame rotating with each nanostructure, which results in
\begin{align}
\left\langle p_i^{{\rm fl,} \pm *}\left(\omega'\right) p_j^{{\rm fl,} \pm}(\omega)\right\rangle = {}& 4 \pi \hbar \delta(\omega-\omega'){\rm Im}\{\chi_i^{\pm}(\omega)\} \delta_{ij}\left[n\left(T_i, \omega \mp \Omega_i \right)+\frac{1}{2}\right] ,
\nonumber \\
\left\langle p_i^{{\rm fl,} z *}\left(\omega'\right) p_j^{{\rm fl,} z}(\omega)\right\rangle = {}& 4 \pi \hbar \delta\left(\omega-\omega'\right) {\rm Im}\{\chi_i^{z}(\omega)\} \delta_{ij}\left[n\left(T_i,\omega\right)+\frac{1}{2}\right] \nonumber,
\end{align}
while, for the fluctuations of the electric field, we have
\begin{equation}
\begin{aligned}
\left\langle E_i^{{\rm fl,} \nu *}\left(\omega'\right) E_j^{{\rm fl,} \nu}(\omega)\right\rangle= 4 \pi \hbar \delta\left(\omega-\omega'\right) {\rm Im}\left\{\delta_{ij} g_0(\omega)+(1 - \delta_{ij})g^{\nu}(\omega)\right\}\left[n(T_0,\omega)+\frac{1}{2}\right], \nonumber
\end{aligned}
\end{equation}
where $n(T_i,\omega)=\left[\exp \left(\hbar \omega / k_{\rm B} T_i\right)-1\right]^{-1}$ is the Bose-Einstein distribution and $\chi_i^{\nu}(\omega)=\alpha_{i}^{\nu}(\omega)- g_0 \left|\alpha_{i}^{\nu}(\omega)\right|^2$. Notice that the Bose-Einstein distribution that appears in the fluctuations of the $\pm$ components of the dipole is frequency-shifted due to the rotation of the nanostructures. On the contrary, the fluctuations of the electric field are associated with the environment, and therefore are not affected by the rotation of the nanostructures. Once the average over fluctuations is performed, the torque exerted on nanostructure $1$ is given by
\begin{equation}
M_1=-\int_{0}^{\infty} d\omega \left\{ \left[F^{+}(\omega) N_1^{-}(\omega) - G^{+}(\omega) N_2^{-}(\omega)\right] - \left[F^{-}(\omega)N_1^{+}(\omega) - G^{-}(\omega)N_2^{+}(\omega)\right] \right\},  \label{SI_M2}
\end{equation}
while the power radiated reads
\begin{align}
P_1={}&\int_{0}^{\infty} d\omega \omega\left\{\left[F^{+}(\omega)N_1^{-}(\omega)-G^{+}(\omega)N_2^{-}(\omega)\right] + \left[F^{-}(\omega)N_1^{+}(\omega)-G^{-}(\omega)N_2^{+}(\omega)\right]\right\} \nonumber \\
&+2\int_{0}^{\infty} d\omega \omega\left[F^z(\omega)N_1^z(\omega)-G^z(\omega)N_2^z(\omega)\right]. \label{SI_P2}
\end{align}
Here, $N_i^{\pm}(\omega)=n\left(T_i, \omega \pm \Omega_i\right)-n\left(T_0, \omega\right) $, $N_i^{z}(\omega) = n\left(T_i, \omega\right)/2-n\left(T_0, \omega\right)/2 $, while the functions $F^\nu(\omega)$ and $G^\nu(\omega)$ are defined as
\begin{align}
F^\nu(\omega)= {}& \frac{2 \hbar}{\pi}\left|h^\nu(\omega)\right|^2 {\rm Im}\{\chi_1^\nu (\omega)\} {\rm Im}\left\{g_0(\omega)+\alpha_2^\nu(\omega)\left(g^\nu(\omega)\right)^2\right\}, \nonumber \\
G^\nu(\omega)= {}& \frac{2 \hbar}{\pi}\left|h^\nu(\omega) \right|^2 \left|g^\nu (\omega) \right|^2 {\rm Im}\{\chi_1^\nu (\omega)\} {\rm Im}\{\chi_2^\nu (\omega)\}. \nonumber
\end{align}
Equations (\ref{SI_M2}) and (\ref{SI_P2}) are equivalent to Equations~(\ref{M1}) and (\ref{P1}) from the main text. Notice that a similar derivation can be performed for the torque and the power radiated associated with the magnetic polarization. However, this contribution is negligible for very small nanostructures in which the magnetic polarizability is generally small, although it can play a role for large nanostructures with high values of conductivity \cite{ama9,ama20}. 

Upon application of the approximations described in the main text, the functions $F^{\pm}(\omega)$ and $G^{\pm}(\omega)$ reduce to 
\begin{equation}
F^{\pm}(\omega)= G^{\pm}(\omega) \approx \frac{2 \hbar}{\pi}\frac{1}{l^6}{\rm Im}\{\alpha_1^{\pm}(\omega)\} {\rm Im}\left\{\alpha_2^{\pm}(\omega)\right\}, \nonumber
\end{equation}
while $\alpha_i^{\pm}(\omega) \approx a_i [1 + i (\omega \mp \Omega_i) \gamma_i / \omega_{{\rm r}, i}^2]$ (see next section). Note that, with these approximations, the term of $F^{\pm}(\omega)$ proportional to $g_0$, which describes the transfer to the environment via far-field radiation, is neglected. Furthermore, the expression for the torque reduces to
\begin{equation}
\begin{aligned}
M_1 = {} & -12 C \int_0^{\infty} d \omega(\omega-\Omega_1)(\omega-\Omega_2)\left[n(T_1, \omega-\Omega_1)-n(T_2, \omega-\Omega_2)\right] \\
& +12 C \int_0^{\infty} d \omega(\omega+\Omega_1)(\omega+\Omega_2)\left[n(T_1, \omega+\Omega_1)-n(T_2, \omega+\Omega_2)\right], \nonumber
\end{aligned}
\end{equation}
while the power transferred reads
\begin{equation}
\begin{aligned}
P_1 = {} & 12 C \int_0^{\infty} d \omega \omega(\omega-\Omega_1)(\omega-\Omega_2)\left[n(T_1, \omega-\Omega_1)-n(T_2, \omega-\Omega_2)\right] \\
& +12 C \int_0^{\infty} d \omega \omega(\omega+\Omega_1)(\omega+\Omega_2)\left[n(T_1, \omega+\Omega_1)-n(T_2, \omega+\Omega_2)\right], \nonumber
\end{aligned}
\end{equation}
where 
\begin{equation}
C = \frac{\hbar}{6 \pi}\left(\frac{\gamma_{1} \gamma_{2}}{\omega_{\rm r, 1}^2 \omega_{\rm r, 2}^2}\right) \left(\frac{a_1 a_2}{l^6}\right). \nonumber
\end{equation}
In order to perform the integrals over frequencies appearing in these expressions, we follow three steps: (1) Separate the terms involving different Bose-Einstein distributions into different integrals; (2) For each integral, apply the change of variables $\omega^{\prime} = \omega \pm \Omega_i$, thus changing the integration limits to $(\pm \Omega_i, \infty)$; (3) Split each resulting integral into one integral over $(0, \infty)$ and another over $(\pm \Omega_i, 0)$, both of which have an analytical solution, and compute them. Following these steps, we readily obtain Equations~(\ref{M1_a}) and (\ref{P1_a}) of the main text.

\subsection{Effective polarizability of a rotating axially symmetric nanostructure}

We can model the polarizability of an axially symmetric nanostructure, produced by an electromagnetic resonance, using a harmonic oscillator model. We assume that the axis of symmetry of the nanostructure corresponds to the $z$ axis and neglect its  response along that axis. Under these conditions, the motion of the charges that gives rise to the electromagnetic resonance obeys the following equation:
\begin{equation}
\ddot{\bf r}(t)=-\omega_{\rm r}^2 {\bf r}(t)-\gamma \dot{\bf r}(t) +\frac{Q}{M} \mathbf{E}(t). \nonumber
\end{equation}
Here, ${\bf r}$ is the position vector of the center of mass of the charges in the $xy$ plane, $\omega_{\rm r}$ is the resonance frequency, $\gamma$ is the nonradiative damping rate, $Q$ and $M$ are, respectively, the total charge and mass of the charges involved in the electromagnetic resonance, and ${\bf E}$ is an external electric field. Note that we have neglected the Abraham-Lorentz term describing the radiative losses \cite{ama10}, in consistency with the approximations described in the main text. 

When the nanostructure is set to rotate around the $z$ axis with rotation frequency $\Omega$, the nonradiative damping term needs to be modified to properly account for the relative velocity between the charges that give rise to the resonance and the ionic background of the nanostructure \cite{PXG21}
\begin{equation}
\ddot{\bf r}(t)=-\omega_{\rm r}^2 {\bf r}(t)-\gamma [\dot{\bf r}(t) - {\bf \Omega} \times  {\bf r}(t)] +\frac{Q}{M} \mathbf{E}(t), \nonumber
\end{equation}
where ${\bf \Omega} = \Omega {\hat{\bf z}}$. It is worth noting that it has been shown \cite{PXG19_2} that the rotation can induce changes in the electronic distribution of the nanostructure that lead to a modification of its resonance frequency for certain geometries. However, this effect, which is beyond the simple harmonic oscillator model, does not play a relevant role in our case because we assume that the rotation frequency is much smaller than the resonance frequency. 

To explore the effect of the noninertial forces, we transform the equation of motion to a reference frame rotating with the nanostructure using the following relations:
\begin{equation}
{\bf r}={\tilde{\mathbf{r}}}, \quad \dot{\mathbf{r}}=\dot{\tilde{\mathbf{r}}}+ {\bf \Omega} \times \tilde{\bf r}, \quad \ddot{\bf r}=\ddot{\tilde{\mathbf{r}}}+2 {\bf \Omega} \times \dot{\tilde{\mathbf{r}}}-\Omega^2 {\tilde{\mathbf{r}}}, \nonumber
\end{equation} 
where the $^{\sim}$ denotes the variables in the rotating frame. By doing so, the equation of motion becomes
\begin{equation}
\ddot{\tilde{\mathbf{r}}}(t) = -\omega_{\rm r}^2 {\tilde{\mathbf{r}}}(t)-\gamma \dot{\tilde{\mathbf{r}}}(t) - 2 {\bf \Omega} \times \dot{\tilde{\mathbf{r}}}(t) + \Omega^2 {\tilde{\mathbf{r}}} (t)  +\frac{Q}{M} \tilde{\mathbf{E}}(t). \nonumber
\end{equation} 
The new terms $- 2 {\bf \Omega} \times \dot{\tilde{\mathbf{r}}}(t)$ and $\Omega^2 {\tilde{\mathbf{r}}} (t)$ correspond to the Coriolis and the centrifugal acceleration, respectively. Projecting into a circular basis defined as $\hat{\tilde{ \mathbf{e}}}^{\pm} =( \hat{\tilde{ \mathbf{x}}} \pm i \hat{\tilde{ \mathbf{y}}})/\sqrt{2}$, we get
\begin{equation}
\ddot{\tilde{r}}^{\pm}(t)=-\omega_{\rm r}^2 {\tilde{r}}^{\pm}(t)-\gamma \dot{\tilde{r}}^{\pm}(t) \pm 2i \Omega \dot{\tilde{r}}^{\pm}(t) + \Omega^2 {\tilde{r}}^{\pm}(t) +\frac{Q}{M} \tilde{E}^{\pm}(t). \nonumber
\end{equation} 
Assuming that $\tilde{E}^{\pm}(t)$, and consequently $\tilde{r}^{\pm}(t)$, oscillate at frequency $\omega$, the polarizability of the nanostructure in the rotating frame is given by
\begin{equation}
\tilde{\alpha}^{\pm}(\omega)= \frac{Q \tilde{r}^{\pm}(\omega)}{\tilde{E}^{\pm}(\omega)} =\frac{a \omega_{\rm r}^2}{\omega_{{\rm r}}^2-(\omega \pm \Omega)^2-i \gamma \omega}, \nonumber
\end{equation}
where we have defined $a = Q^2 / (M \omega_{\rm r}^2)$. Transforming this polarizability back to the frame at rest, we obtain the following effective polarizability:
\begin{equation}
\alpha^{\pm}(\omega) = \tilde{\alpha}^{\pm}(\omega \mp \Omega) = \frac{a \omega_{\rm r}^2}{ \omega_{\rm r}^2 -\omega^2 -i \gamma (\omega \mp \Omega)}.\nonumber
\end{equation} 
Notice that, under the conditions considered here, in which we neglect the radiative losses, the rotation frequency only enters the effective polarizability through the nonradiative damping term.


\end{document}